\documentclass{article}
\usepackage{preprint}
\usepackage{epsf}
\usepackage{psfig}
\def\FUSE{{\it FUSE}}
\def\apj{{\it Ap.J.}}
\def\apjs{{\it Ap.J.Supp.}}
\def\aj{{\it A.J.}}
\def\mnras{{\it M.N.R.A.S.}}
\def\arcsec{\ifmmode '' \else $''$\fi}
\def\arcmin{\ifmmode ' \else $'$\fi}
\def\arcsecpoint{\ifmmode ''\!. \else $''\!.$\fi}
\def\arcminpoint{\ifmmode '\!. \else $'\!.$\fi}
\def\cc{\ifmmode {\rm cm}^{-3} \else cm$^{-3}$\fi}
\def\cl{\ifmmode {\rm cm}^{-2} \else cm$^{-2}$\fi}
\def\micron{\ifmmode \mu{\rm m} \else $\mu$m\fi}
\def\kms{\ifmmode {\rm km\,s}^{-1} \else km\,s$^{-1}$\fi}
\def\Hubble{\ifmmode {\rm km\,s}^{-1}\,{\rm Mpc}^{-1}
        \else km\,s$^{-1}$\,Mpc$^{-1}$\fi}
\def\ergsec{\ifmmode {\rm ergs\;s}^{-1} \else ergs s$^{-1}$\fi}
\def\ergscm{\ifmmode {\rm ergs\,s}^{-1}\,{\rm cm}^{-2}
          \else ergs\,s$^{-1}$\,cm$^{-2}$\fi}
\def\ergscmA{\ifmmode {\rm ergs\,s}^{-1}\,{\rm cm}^{-2}\,{\rm \AA}^{-1}
          \else ergs\,s$^{-1}$\,cm$^{-2}$\,\AA$^{-1}$\fi}
\def\ergscmHz{\ifmmode {\rm ergs\,s}^{-1}\,{\rm cm}^{-2}\,{\rm Hz}^{-1}
          \else ergs\,s$^{-1}$\,cm$^{-2}$\,Hz$^{-1}$\fi}
\def\Msun{\ifmmode M_{\odot} \else $M_{\odot}$\fi}
\def\Lsun{\ifmmode L_{\odot} \else $L_{\odot}$\fi}
\def\qo{\ifmmode q_{0} \else $q_{0}$\fi}
\def\Ho{\ifmmode H_{0} \else $H_{0}$\fi}
\def\ciii{C\,{\sc iii}}
\def\civ{C\,{\sc iv}}

\newcommand{\ovi}{O~{\sc vi}}

\newcommand{\heii}{He~{\sc ii}}

\newcommand{\siv}{S~{\sc iv}}

\def\eps@scaling{.95}
\def\epsscale#1{\gdef\eps@scaling{#1}}
\def\plotone#1{\centering \leavevmode
\epsfxsize=\eps@scaling\columnwidth \epsfbox{#1}}

\begin{document}

\title{
A Far-Ultraviolet Spectroscopic Survey of Low-Redshift AGN}

\author{Gerard~A. KRISS and the FUSE AGN Working Group}


\affil{Space Telescope Science Institute\\3700 San Martin Drive\\Baltimore, MD 21218\\gak@stsci.edu}



\begin{abstract}
Using the Far Ultraviolet Spectroscopic Explorer (FUSE) we have obtained 87
spectra of 57 low-redshift ($z<0.15$) active galactic nuclei (AGN). This sample
comprises 53 Type 1 AGN and 4 Type 2. All the Type 1 objects show broad
\ovi\ $\lambda 1034$ emission;
two of the Type 2s show narrow \ovi\ emission. In addition
to \ovi, we also identify emission lines
due to \ciii\ $\lambda 977$, \ciii\ $\lambda 991$,
\siv\ $\lambda\lambda 1062,1072$, and \heii\ $\lambda 1085$
in many of the Type-1 AGN.
Of the Type 1 objects, 30 show intrinsic absorption by the
\ovi\ $\lambda\lambda 1032,1038$ doublet.
Most of these intrinsic absorption systems show multiple components with
intrinsic widths of 100 \kms\ spread over a blue-shifted velocity
range of less than 1000 \kms.
Galaxies in our sample with existing X-ray or longer wavelength UV observations
also show \civ\ absorption and evidence of a soft X-ray warm absorber.
In some cases, a UV absorption component has physical properties similar to the
X-ray absorbing gas, but in others there is no clear physical correspondence
between the UV and X-ray absorbing components. Models in which a thermally
driven wind evaporates material from the obscuring torus naturally produce such
inhomogeneous flows.
\end{abstract}

\section{Introduction}

Roughly 50\% of all Seyfert galaxies show UV absorption lines, most
commonly seen in {\sc C~iv} and Ly$\alpha$ \citep{Crenshaw99}.
X-ray ``warm absorbers" are equally common in
Seyferts \citep{Reynolds97, George98}.
All instances of X-ray absorption
also exhibit UV absorption \citep{Crenshaw99}.
While \citet{Mathur94} and \citet{Mathur95} have suggested that the same
gas gives rise to both the X-ray and UV absorption, the spectral complexity
of the UV and X-ray absorbers indicates that a wide range of physical
conditions are present.
Multiple kinematic components with differing physical conditions are seen
in both the UV \citep{Crenshaw99, Kriss00b} and in the
X-ray \citep{Reynolds97, Kriss96, Kaspi01}.

The short wavelength response (912--1187 \AA) of the
{\it Far Ultraviolet Spectroscopic Explorer (FUSE)} \citep{Moos00}
enables us to make high-resolution spectral measurements
($R \sim 20,000$) of the high-ionization
ion {\sc O~vi} and the high-order Lyman lines of neutral hydrogen.
The \ovi\ doublet is a crucial link for establishing a connection between
the higher ionization absorption edges seen in the X-ray and the lower
ionization absorption lines seen in earlier UV observations.
The high-order Lyman lines provide a better constraint on the
total neutral hydrogen column density than Ly$\alpha$ alone.
Lower ionization species such as {\sc C~iii} and {\sc N~iii} also have strong
resonance lines in the \FUSE\ band, and these often are useful for setting
constraints on the ionization level of any detected absorption.
The Lyman and Werner bands of molecular hydrogen also fall in the \FUSE\ band,
and we have searched for intrinsic $\rm H_2$ absorption that may be
associated with the obscuring torus.

We have been conducting a survey of the $\sim100$ brightest AGN using
{\it FUSE}. As of November 1, 2002, we have observed a total of 87;
of these, 57 have $z < 0.15$, so that the {\sc O~vi} doublet is visible
in the FUSE band.

\section{Survey Results}

Over 50\% (30 of 53) of the low-redshift Type 1 AGN observed using \FUSE\ show
detectable \ovi\ absorption,
comparable to those Seyferts that show
longer-wavelength UV \citep{Crenshaw99}
or X-ray \citep{Reynolds97, George98} absorption.
None show $\rm H_2$ absorption.
We see three basic morphologies for \ovi\ absorption lines:
(1) {\bf Blend}: multiple \ovi\ absorption components that are blended
together.  
10 of 30 objects fall in this class, and the spectrum of Mrk~509
is typical \citep{Kriss00b}.
(2) {\bf Single}: 13 of 30 objects exhibit single, narrow,
isolated \ovi\ absorption lines, as
illustrated by the spectrum of Ton~S180 \citep{Turner01}.
(3) {\bf Smooth}: The 7 objects here are an extreme expression of the
``blend" class,
where the \ovi\ absorption is so broad and blended that individual
\ovi\ components cannot be identified.  NGC~4151 typifies
this class \citep{Kriss92}.
Individual \ovi\ absorption components have
FWHM of 50--750 \kms, with most objects having FWHM $< 100~\kms$.
The multiple components that are typically present are almost always
blue shifted, and they span a velocity range of 200--4000 \kms;
half the objects span a range of $< 1000~\kms$.

\section{Discussion}

The multiple kinematic components frequently seen in the UV absorption spectra
of AGN clearly show that the absorbing medium is complex, with separate
UV and X-ray dominant zones.
In some cases, the UV absorption component corresponding to the X-ray warm
absorber can be clearly identified (e.g., Mrk~509) \citep{Kriss00b}.
In others, however, {\it no} UV absorption component shows physical
conditions characteristic of those seen in the X-ray absorber
(NGC~3516, NGC~5548) \citep{Kriss96, Brotherton02}.
One potential geometry for this complex absorbing structure is high-density,
low-column UV-absorbing clouds embedded in a low-density,
high-ionization medium that dominates the X-ray absorption.
This is possibly a wind driven off the obscuring
torus \citep{KK95, KK01}.
At the critical ionization parameter for evaporation, there is a broad range of
temperatures that can coexist in equilibrium at nearly constant pressure;
for this reason, the flow is expected to be strongly inhomogeneous.
What would this look like in reality?
As a nearby analogy, consider 
the HST images of the pillars of gas in the Eagle Nebula, M16. These
show the wealth of detailed structure in gas evaporated from a molecular
cloud by the UV radiation of nearby newly formed stars \citep{Hester96}.
Figure 1 shows what this might look like in an AGN---a dense molecular torus
surrounded by blobs, wisps, and filaments of gas at various densities.
It is plausible that the multiple UV absorption lines seen in AGN with
warm absorbers are caused by high-density blobs of gas embedded in a hotter,
more tenuous, surrounding medium, which is itself responsible for the X-ray
absorption.  Higher density blobs would have lower ionization
parameters, and their small size would account for the low overall column
densities.

\begin{figure}[ht]
\centerline{\epsfxsize=3.7in\epsfbox{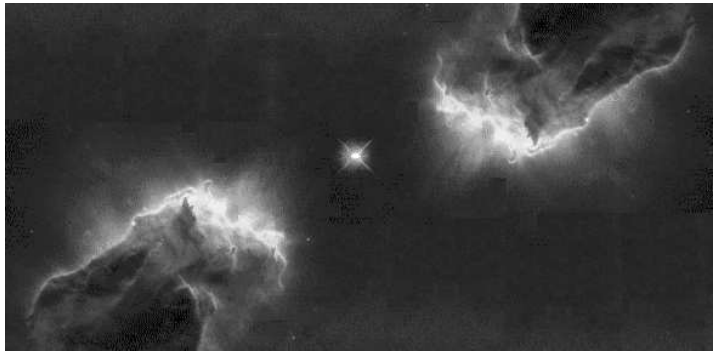}}   
\caption{
An artist's conception of how a molecular torus surrounding an AGN might appear
based on HST images of the Eagle Nebula. Note the complex of wisps and blobs
of gas close to the surface of the molecular material.
\label{torus}}
\end{figure}

At sight lines close to the surface of the obscuring torus, one might
expect to see some absorption due to molecular hydrogen.
Given the dominance of Type 1 AGN in our observations so far, the lack of
any intrinsic $\rm H_2$ absorption is not too surprising since our
sight lines are probably far above the obscuring torus.
NGC~4151 and NGC~3516 are examples where the inclination may be more favorable
since these objects have shown optically thick Lyman limits in the
past \citep{Kriss92, Kriss96},
but our
FUSE observations do not show such high levels of neutral hydrogen.
Molecular hydrogen will not survive long in an environment with a strong
UV flux, and this probably accounts for the lack of $\rm H_2$ absorption.

In summary, we find that \ovi\ absorption is common in low-redshift ($z < 0.15$)
AGN. 30 of 53 Type 1 AGN with $z < 0.15$ observed using \FUSE\ show
multiple, blended \ovi\ absorption lines with typical widths
of $\sim 100~\kms$ that are blueshifted over a velocity range of $\sim$
1000 \kms.
Those galaxies in our sample with existing X-ray or longer wavelength UV
observations also show {\sc C~iv} absorption and evidence of a soft X-ray
warm absorber.
In some cases, a UV absorption component has physical properties
similar to the X-ray absorbing gas, but in others there is no clear
physical correspondence between the UV and X-ray absorbing components.

\section*{Acknowledgments}
This work is based on data obtained for the Guaranteed Time Team by the
NASA-CNES-CSA FUSE mission operated by the Johns Hopkins University. Financial
support to U. S. participants has been provided by NASA contract NAS5-32985.

\end{document}